%% file: ICRC2025_Ice.tex
\documentclass[a4paper,11pt]{article}

\usepackage{pos}
\usepackage{lipsum} 
\usepackage{wrapfig}

\title{State of the Ice Model in the IceCube Observatory}

\ShortTitle{Ice Model}

\author{The IceCube Collaboration \\{\normalsize \normalfont(a complete list of authors can be found at the end of the proceedings)}\\}

\emailAdd{dima@icecube.wisc.edu}
\emailAdd{martin.rongen@fau.de}

\abstract{

IceCube is a neutrino observatory located at the South Pole that uses Antarctic ice as a medium for detection of Cherenkov photons. As such, analysis of the data relies on our understanding of the properties of ice within and around the instrumented volume. Over the years we have made significant progress in understanding the glacial ice and now have a comprehensive model that covers many of the relevant aspects of the photon propagation in it. In this report we give a historical overview of the ice description within the IceCube detector, list some of the remaining issues, and assess how much more improvement is still needed. As the IceCube Upgrade is expected to be installed in less than a year, with several new types of calibration devices aiming to further our understanding of ice, this is the perfect time to review the current state of the ice model.

\vspace{4mm}

{\bfseries Corresponding authors:}
Dmitry Chirkin$^{1*}$, 
Martin Rongen$^{2}$\\
{$^{1}$ \itshape Dept. of Physics and Wisconsin IceCube Particle Astrophysics Center, University of Wisconsin{\textendash}Madison, Madison, WI 53706, USA}\\
{$^{2}$ \itshape  Erlangen Centre for Astroparticle Physics, Friedrich-Alexander-Universität Erlangen-Nürnberg}\\
$^*$ Presenter
}

\input{ICRCdetails.tex}

\begin{document}

\maketitle

\section{History of IceCube Ice Models}\label{sec1}

IceCube is a kilometer-cube-scale neutrino detector deployed in optically transparent South Pole ice at depths between 1450 and 2450 m. It consists of 5160 digital optical modules (DOMs), each containing a large 10 inch PMT, 12 LEDs (most are 400 nm), and associated electronics. The DOMs are organized on 86 strings (60 DOMs per string), with 78 of them arranged in a hexagonal pattern, 125 m between neighboring strings, and the rest (8 strings) deployed more densely as part of the DeepCore. IceCube, together with a surface component IceTop, was completed in 2010.

IceCube has been preceded by a smaller AMANDA-II array (Antarctic Muon And Neutrino Detector Array), which consisted of 677 optical modules organized on 19 strings, most deployed between depths of 1500 and 2000 m, but with three of the strings covering a much larger depth range between 1150 and 2350 m. AMANDA-II has been completed in 2000 and eventually decommissioned in 2009.

There was an even earlier array, AMANDA-A, which included 4 strings of optical sensors, deployed at depths between 810 and 1000 m. AMANDA-A and AMANDA-II contained a variety of light sources for calibrations, both continuous and pulsed, some deployed in-situ and some fed from the surface to diffuser balls, including LED "flashers", and nitrogen and YAG lasers.

In 1994 AMANDA-A collaboration published \cite{amanda94} a measurement result of optical properties at depths between 800 and 1000 m, with fairly low absorption with absorption length of 59 m\footnote{at 515 nm, later shown to reach even higher values of over 200 m at lower, more relevant wavelengths (415 nm) \cite{amanda06}}, and very high scattering with {\it effective} scattering length\footnote{average propagation length after which photon direction changes by $\sim \pi/2$; {\it geometric} scattering length is smaller} of around 1 m. This was consistent with the air bubble concentrations expected from two ice cores collected elsewhere in Antarctica, and an extrapolation suggested that the air bubbles should disappear (fully convert to clathrate hydrates, which are virtually indistinguishable from ice in regards to their refractive index) at depths below 1150 m. This number has later been revised (after including AMANDA-II data) to 1350 m.

The balance between scattering and absorption lengths (the former being much smaller than the latter), and the observed propagation distance (tens of meters, i.e., much larger than the scattering length), meant that the expected rates and time distributions of arriving photons could be accurately calculated within the diffusive approximation regime, resulting in a fully analytical description, without need for detailed time-consuming montecarlo simulations.

In 2006 the AMANDA-II collaboration has published its optical properties paper \cite{amanda06}. Using a variety of light sources the authors presented measurements of scattering and absorption tabulated into the 10 m tall ice layers, together with their wavelength dependence. To this day, including all ice models used by IceCube collaboration, the ice model is defined in slices of 10 m ice layers (at one reference xy location), and the wavelength dependence from the AMANDA-II paper. The IceCube detector contains LED light sources with multiple wavelengths (340, 370, 450, 505, and most are 405 nm), which we had used to verify the wavelength dependence established in \cite{amanda06}. We also continue to use the parameterization of air bubble contribution to scattering derived there.

To highlight one of the measurements of the AMANDA-II optical properties paper, each emitter-receiver pair  in the vicinity of a given 10 m tall ice layer was fitted to a single best (for that pair) value of scattering and absorption. These parameters were then averaged across all contributing pairs for the ice layer. Herein lie a simplification and an approximation of the method: instead of a simulation that includes ice layers above and below, with their own (and often different) scattering and absorption coefficients, info the fit, only one set of scattering and absorption coefficients was ever fitted to each emitter-receiver pair (implicitly assuming that these coefficients are valid throughout the detector volume, for that pair). This lead to a wide range of scattering and absorption coefficients from different emitter-receiver pairs contributing to each ice layer, and an inability of the averaged values to describe all pairs simultaneously. For a while, and even during the first few years of IceCube ice analysis, this method continued to drive the analysis of optical properties of ice. There was an attempt to {\it unfold} the true scattering and absorption coefficients into 10 m ice layers to improve the description of all emitter-receiver pairs by a resulting ice model, with limited success. It resulted in an improvement of ice description at intermediate depths, but fell short in deeper ice, where the ice was much clearer (and, presumably, more neighboring ice layers contributed to each emitter-receiver pair).

At this point (around 2009) the photon propagation was performed by a complicated procedure of tabulating photons into binned tables of parameters (such as relative position of emitters and receivers and initial photon direction) and then sampling from the tables. It has been perceived that the direct photon propagation would not be feasible due to large calculation time. However, with ever-faster computers and emergence of GPUs, direct photon propagation became possible \cite{ppc}. A direct simulation of photon propagation has several important advantages. 1: there are no binning artifacts from the partitioned phase space. 2: one can simulate more complicated phase space. And 3: one can quickly change the underlying ice properties. This lead to an immediate significant improvement of quality of ice description across all depths when used with the in-situ LED IceCube data, which was presented in our paper \cite{mie}.

There were a number of initial simplifications (such as simulating light from 6 horizontal LEDs with an azimuthally-symmetric pattern of light) and optical effects we had yet to discover, which made the perfect modeling of ice impossible. In addition, instead of the usual likelihood where the expectation is modeled with a parametric function, we are comparing data to simulation, which is generated on the spot, with associated statistical uncertainties. This lead to a development of a special comparison function (in the following simply called {\it llh}), which compares unknown expectations from data and simulation with a penalty term (modeled as a log-normal distribution PDF of $\sim 10$\% width that we call "model error"), as described in \cite{llh}.

Eight of the IceCube string drilled holes were surveyed with a dust logger device before the string deployment. This device shone a laser beam of light into the ice and collected returned signal with a PMT, which allowed for a fine vertical scan of the ice properties with a near-millimeter resolution. The narrow features such as volcanic ash layers stand out and allow depth matching of the ice layers between the 8 holes as described in \cite{dustlogger}. This resulted in a first {\it ice layer tilt} model, which was first incorporated into the SPICE\footnote{SPICE is an acronym for South Pole ICE, used within names of direct-fitted IceCube ice models}-2 family of models.

For the next ice model, SPICE Mie, described in \cite{mie}, we found that some flexibility in the scattering function (via a phenomenological two-component model) allowed for a more accurate data description. The fitted scattering function was a better approximation to the prediction of the Mie scattering theory when applied to the radii and density distributions of the 4 main impurity categories (mineral, salt, acid, soot) as measured in the ice cores elsewhere in Antarctica, which added credence to the approach.

In 2013 we presented evidence of optical anisotropy a the South Pole in our report \cite{lea}. We found that the LED light propagating along the axis nearly collinear with the direction of the ice flow at the South Pole (which moves at a rate of about 10 m/year in the direction 135 degrees grid North), delivers roughly twice the amount of charge at one string distance (125 m) away, compared to directions perpendicular to it. We have initially developed a phenomenological model, which introduced anisotropy into the scattering function. The effect resulted in less scattering along the anisotropy axis compared to other directions. One could visualize such an effect emerges if the impurities in the ice have an elongated shape (approximated with a prolate spheroid), with long axis of the shape preferentially aligning along the anisotropy axis. This presents a smaller cross section to photons propagating along the anisotropy axis, thus reducing the effective scattering coefficient for those directions. We named the resulting model SPICE Lea, Lea being an acronym for $\lambda_e$\footnote{$\lambda$ is commonly used to denote the scattering length, with subindex $e$ being short for {\it effective}.}-ani.

It was soon discovered that, while improving the description of the time-integrated charge in all receivers surrounding the emitter LEDs, the scattering-based anisotropy failed to describe well the timing structure of the received photons. The peak of the time distributions for directions along the anisotropy axis was pushed into earlier times (as one would expect from less scattering along those directions) compared to data. In fact, the timing structure in data was best described by a simulation without any anisotropy at all. This lead to a renewed effort in studying the possible reasons for the effect. One new possibility included absorption-based anisotropy, resulting in SPICE EMRM (Enhanced Modified Renormalized Martin) ice model. Eventually, both scattering and absorption-based models were ruled out in favor of a new, birefringence-based anisotropy model, which yielded a much improved description of both amplitude and time structure of the received photon signal, especially when mixed in with small amounts of scattering and absorption anisotropies. The effect relies on birefringent refraction and reflection on the surfaces of millimeter-sized ice crystals, which form the bulk of the South Pole ice at detector depths. The crystals gain preferential orientations correlated with the forces related to the ice flow and hydrostatic pressure, which result in the macroscopic effect we observe in IceCube. The photons sustain more (birefringent) scattering along the anisotropy axis, which is counter to the previous descriptions, but also are subject to a slow bending of their paths towards the axis in between larger scatters. The resulting model, analysis and even crystal size depth dependence (extracted from our optical calibration data) are published in \cite{bfr}.

Finally, at the last ICRC we reported on our effort to map the ice layer undulations (tilt) within the volume of the detector \cite{ftp}. We started by investigating local depth shifts to the fitted ice layer depth table that would result in the best description of individual flasher DOMs. Eventually a single solution that best describes all flasher DOMs simultaneously was found. It consists of 9875 ice layer elevation values (tabulated on a hexagonal grid of 80 xy locations approximately matching the outline of the detector with 125 depth values each) and a depth table of scattering, absorption, and ice crystal density values in each ice layer with a total number of fit parameters at over 10300. After two years of running and refining the fit, we produced simulations of a few thousand of ice variations. Together with an iterative paraboloid fitter, purpose-designed for this study, with a number of regularizations and correlation matrix sparsity constraints, this was sufficient to produce a solution together with a full covariance (uncertainty) matrix, which we named SPICE FTP (Full-Tilt Parameterization).

\section{Ancillary parameters}\label{sec2}

Each IceCube DOM contains 12 LEDs, usually 6 horizontal and 6 tilted (elevated 45 degrees up from horizontal), arranged in a regular pattern around the circle (60 degrees apart from each other). The first flasher dataset was taken around a single IceCube string before the detector was completed, with data from six more strings added in 2010. The {\it all-purpose} set was taken a couple years later with a completed detector and collected data from every working DOM on 85 strings. These sets used 6 horizontal or 6 tilted DOMs simultaneously, creating a pattern of light that was initially approximated as cylindrically-symmetric in simulation. For the {\it single-LED} dataset taken in 2018, which flashed each of the LEDs individually, we fitted azimuthal orientations of DOMs (most with better than 1 degree uncertainty) to accurately model light from each LED.

The cable supporting DOMs on each string is routed around each DOM, nominally in the same place between LEDs 11 and 12. It casts a shadow on a DOM PMT, therefore its position can be reconstructed, albeit to $\sim$ 60 degree uncertainty (which was verified with a correlation to the azimuthal orientation of the DOMs mentioned above). We usually take the nominal routed position of the cable with respect to the DOMs azimuthally-rotated with a more precise method above, which we refer to as the emitter-side reconstruction.

Although nominally vertical, DOMs may freeze-in with a small tilt of their PMT axis from the vertical. The receiver-side reconstruction of the tilt has been previously reported but may mimic the effect from the cable shadow or hole ice (discussed below). Nevertheless, we have observed a correlation for DOMs with largest fitted tilt and readings from the inclinometer boards (with which 48 of the DOMs are equipped). The more precise, but also more time-consuming, emitter-side fits have also been performed recently and show a more robust level of this correlation.

Relative in-ice DOM Efficiencies (RDEs) have been re-fitted on a regular basis, and most recently (fitted with the SPICE FTPv3 ice model) the fitted values showed visible correlation to the lab-measured values, which is something we had not seen with the earlier ice models.

Two camera modules (Sweden cameras) were installed at the bottom of the last deployed IceCube string. They were able to shed light on the structure of ice after the water in the hole freezes. The freezing starts from the outer edges of the hole and pushes bubbles and other impurities inwards, leaving very clear ice behind the ice-water interface. Eventually the bubbles start freezing in place, forming a narrow column of highly-scattering bubbly ice in the center of the hole with a diameter about a third of the initial hole diameter. The DOM diameter being smaller than the hole (around two thirds), the DOM may freeze anywhere in the hole (although we hypothesize the most likely situation, and the one observed by the Sweden cameras, is where the DOM touches the hole wall and stays there after freezing). We have previously simulated the effect of such hole ice structure at both the emitter and receiver sides, and fitted the effective scattering length inside the bubbly column to $\sim$ 3 cm, and the relative positions of DOMs with respect to the bubbly column. The emitter-side fit yields a much more constraining result as the light that initially goes into the bubbly column will result in a dramatically different event compared to the one that goes out into the bulk ice unimpeded.

Effects described in this section have been previously reported in \cite{local}. New efforts in re-fitting the DOM tilt and hole ice properties are currently under way. Recently we have also combined many of the ancillary parameters (i.e., parameters other than the tilt map and depth-dependent table of scattering, absorption and ice crystal density) and attempted to refine their values in a single fit. The parameters included scaling factors for the depth-dependent ice properties, LED elevation angles, ice anisotropy configuration, etc. We were able to further improve the quality of description of the calibration data, but have not yet reached the level of "statistical floor" described in the next section.

\begin{figure}[t!]
\centering
\includegraphics[width=0.45\linewidth]{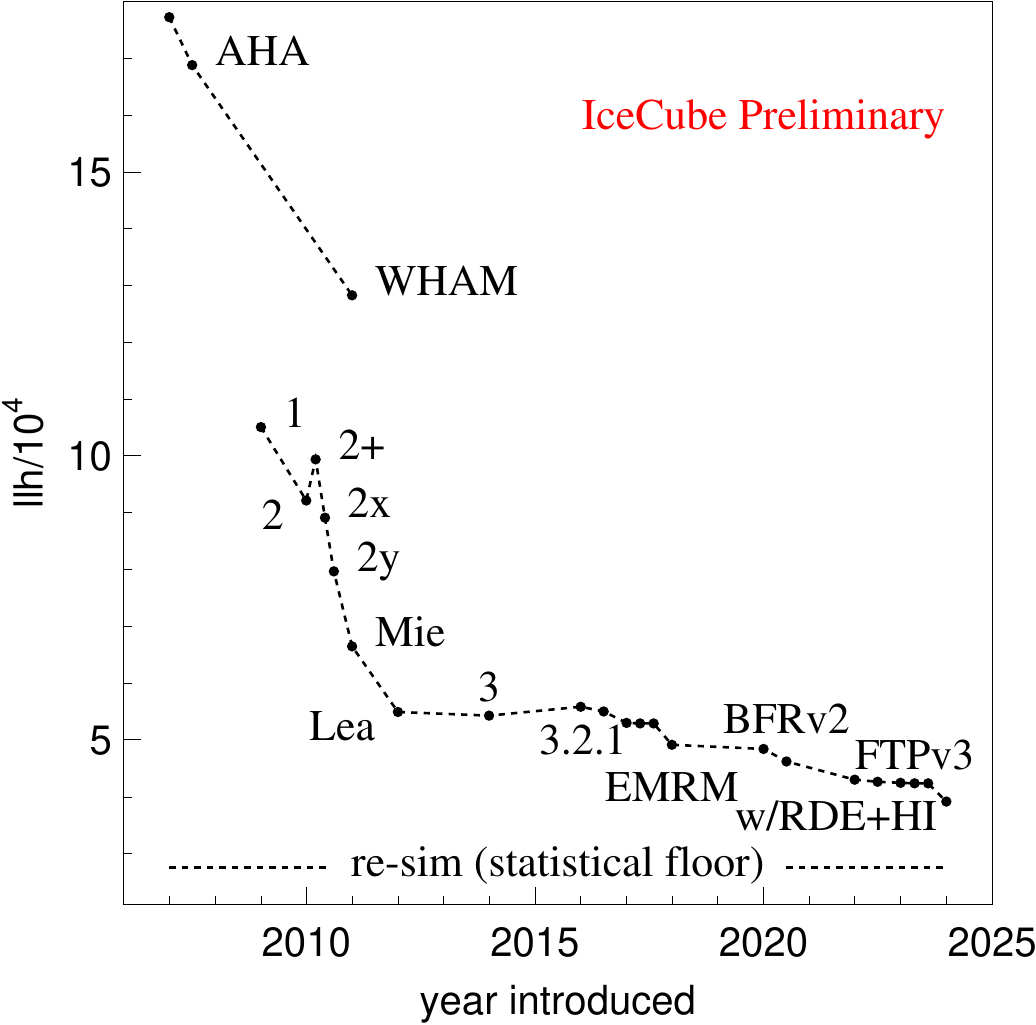}
\includegraphics[width=0.45\linewidth]{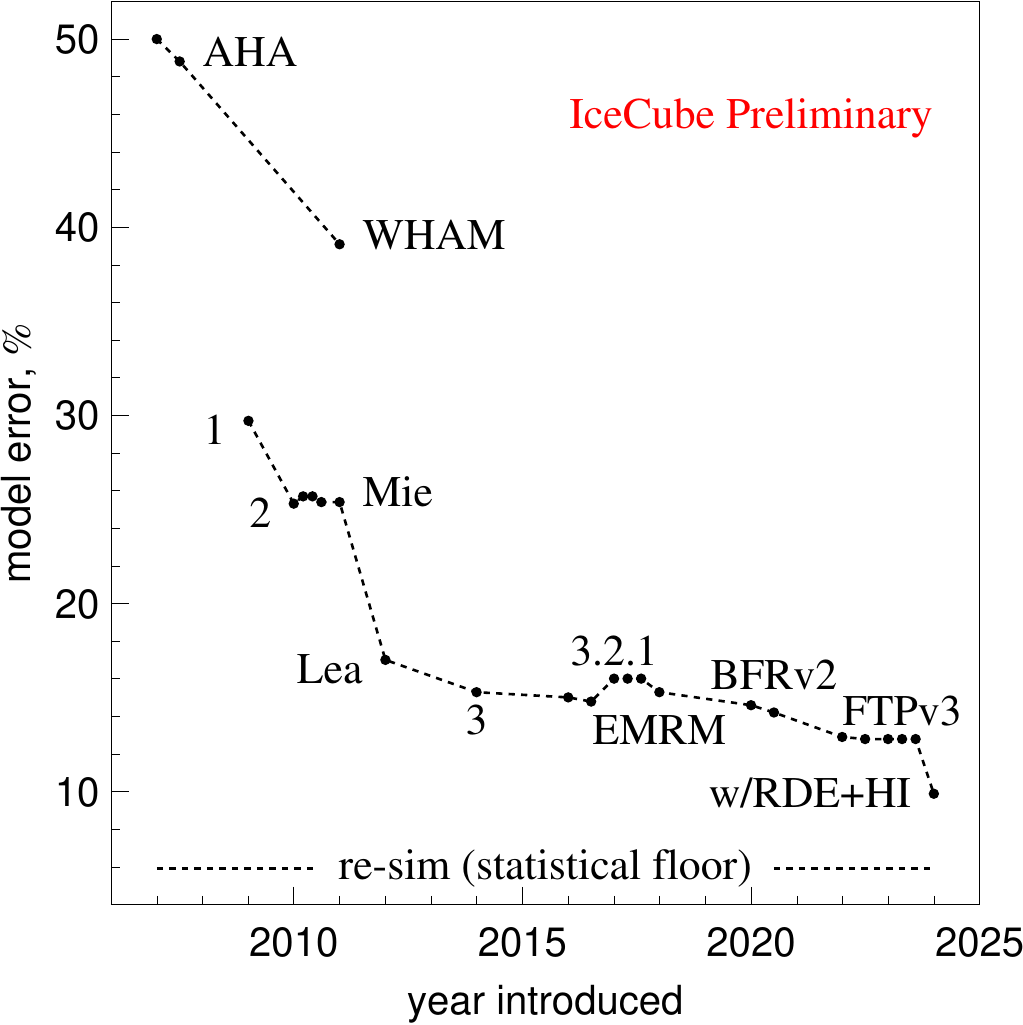}
\caption{Progression of the ice model quality of description of flasher data over the years. {\it llh} is shown on the left, model error on the right. AHA and WHAM are AMANDA-style fits with unfolding, the latter using IceCube calibration data. All other models are "SPICE" (which is omitted from label). In several cases only the last model in a sequence of similar models is labeled (for brevity). Statistical floor is indicated with a dashed line at the bottom of each plot.}\label{llh}
\end{figure}

\section{Visualizing our progress}

Figure \ref{llh} visualizes our progress over the years in quality of description of our flasher LED calibration dataset. The quantity shown as {\it llh}, introduced in section \ref{sec1}, is constructed from two Poisson terms (one for data, one for simulation) and a model error term which ties together the (unknown) expectation values for data and simulation. The quantity is a sum over all ($\sim$ 60k) LED terms, each divided by number of contributing bins and normalized in a way similar to saturated Poisson, which makes {\it llh} a measure of the goodness-of-fit.

The other quantity, shown on the right is the "model error", which is a measure of the difference in time-integrated charges between data and simulation. The two plots show several similarities: There was a large improvement going from the AMANDA-style analysis (which relied on tabulated 1-layer simulation followed by unfolding) to the on-the-spot re-simulation of the entire ice depth table. The next feature is the drop before and flattening after SPICE Lea - this is the ice model that first included ice anisotropy. The last model shown in the plot is SPICE FTP v3, after which we applied several additional fits ancillary to the bulk ice parameters: mainly to the relative DOM efficiencies (RDEs), and optical properties and relative position of the bubbly hole ice column.

\begin{wrapfigure}{r}{0.45\textwidth}
\includegraphics[width=\linewidth]{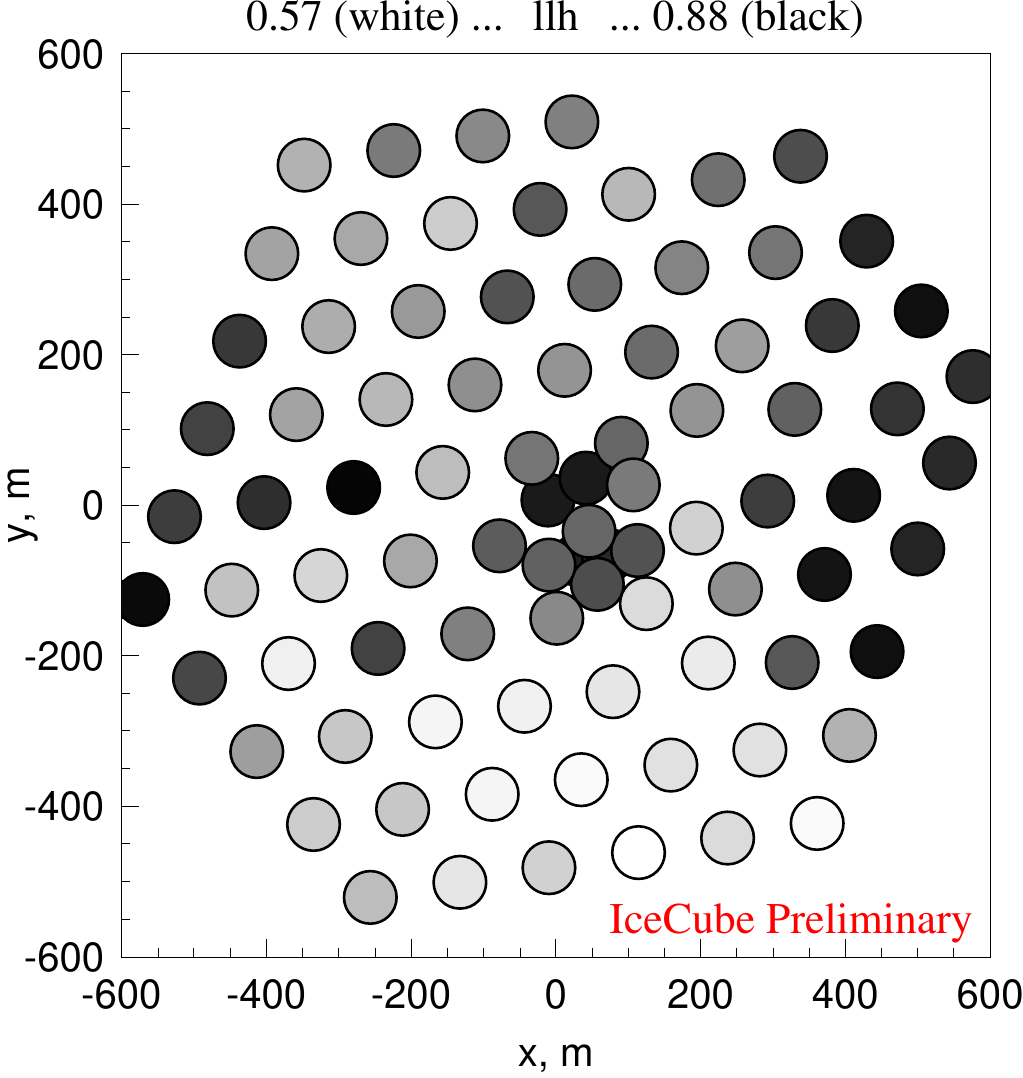}
\caption{Average {\it llh} (over all LEDs) on each string. Overall average for the entire detector is 0.664, compare with that from reconstruction of simulated data: range of 0.42 to 0.45 with the average of 0.439.}\label{str}
\end{wrapfigure}

To gauge how close we are to the perfect ice description, we have simulated a replica of the entire single LED flasher dataset using the best (at the time) fitted ice, RDEs, hole ice parameters, etc. This was reconstructed in the same way as data, and resulted in {\it llh} and model error values shown in Figure \ref{llh} with a label "statistical floor". While we have made great progress over the years, it appears that there is still room for improvement. The current direction of our effort includes refinements in our DOM simulation (including detailed simulation of LED position and obstructions inside DOMs and a possibility of DOM tilts) and fits to the hole ice. We have also looked at possible re-crystallization of the ice in the vicinity of the deployed strings, possibly facilitated by the warming of the ice during drilling and the over-pressures thereafter, as evidenced by spikes in pressure as high as twice the hydrostatic pressure during the hole freezing. We are also continuing with the efforts in the fits to the ancillary parameters, with a possibility that they may vary over the horizontal extent of the ice layers. Figure \ref{str} shows the {\it llh} value averaged per string, and suggests a possibility of horizontal variations in the ice properties (that go beyond ice layer undulations).

We also continue to improve our data analysis even before the data reaches the {\it llh} calculation. We have developed a method for identification of cases of unreliable data, which passed our initial cleaning efforts. It is possible, e.g., that the data acquisition hardware was not ready to start recording a waveform at the start of a flasher event, and instead started late or did not start at all. Such cases can be identified by verifying that the charge recorded in a sequence of same-configuration flasher events follows Poisson distribution. The RMS is predicted from an observed mean of the charges in a sequence, and compared with the actual RMS with a $\chi^2$ distribution. If the probability of observing such an RMS falls below $10^{-5}$, the sequence is rejected. This resulted in a significant improvement in the quality of description of our calibration data, removing many of the outliers.

We have also recently improved our time-binning method, which has been based on the Bayesian block algorithm (referenced in \cite{llh}). We noted that in an overwhelming majority of our waveforms within the fitted range of $\sim$ 1500 ns, the waveform should have only one maximum (ignoring statistical fluctuations), monotonically rise before, and monotonically fall after the maximum. We have modified the Bayesian block algorithm to explicitly take this condition into account. This removed spurious statistical spikes, and reduced the effect from a data feature extraction artifact, where the extracted pulse series may develop oscillating behavior due to small errors in the description of a single-p.e.\ pulse.

\section{The IceCube Upgrade}
After the progress of the past years, there remain no striking data-MC disagreements in the existing LED data that could inform us on how to obtain the large potential gains that are necessary to reach the statistical floor. 
We are thus looking forward to the new wealth of data that will be coming from the IceCube Upgrade \cite{IceCubeUpgrade}, which will be deployed later this year. This in particular includes new propagation baselines due to the different array spacing, a wide range of wavelengths available from the POCAM \cite{POCAM}, directional sweeps with the Pencil Beam \cite{pencilbeam} and visual inspection of local properties via a large number of cameras (either standalone or integrated in the sensor modules \cite{camera}). 

\section{Conclusion}\label{sec3}

We have made much progress in understanding of optical properties of the South Pole ice over the last 30 years. From AMANDA-A discovering air bubbles, to most recently discovering ice anisotropy due to the birefringent nature of ice with IceCube, our description of ice has steadily improved, and we reported on the progress at this conference and in several journal publications. It appears that we are still missing some pieces of the ice puzzle and have not yet reached the "statistical floor" with our description. We are currently evaluating possible further improvements with existing IceCube calibration data, and are looking forward to new data from the IceCube Upgrade, as soon as next year.

\bibliographystyle{ICRC}
\bibliography{references}

%

\clearpage

\input{authorlist_IceCube.tex}

\end{document}

%% file: ICRCdetails.tex
\ConferenceLogo{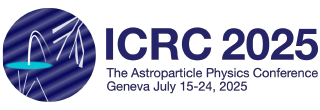}

\FullConference{39th International Cosmic Ray Conference (ICRC2025)\\
 15–24 July 2025\\
Geneva, Switzerland\\}

%% file: authorlist_IceCube.tex
\section*{Full Author List: IceCube Collaboration}

\scriptsize
\noindent
R. Abbasi$^{16}$,
M. Ackermann$^{63}$,
J. Adams$^{17}$,
S. K. Agarwalla$^{39,\: {\rm a}}$,
J. A. Aguilar$^{10}$,
M. Ahlers$^{21}$,
J.M. Alameddine$^{22}$,
S. Ali$^{35}$,
N. M. Amin$^{43}$,
K. Andeen$^{41}$
% ...
\\
\\
% ...
$^{10}$ Universit{\'e} Libre de Bruxelles, Science Faculty CP230, B-1050 Brussels, Belgium \\
% ...
$^{16}$ Department of Physics, Loyola University Chicago, Chicago, IL 60660, USA \\
$^{17}$ Dept. of Physics and Astronomy, University of Canterbury, Private Bag 4800, Christchurch, New Zealand \\
% ...
$^{21}$ Niels Bohr Institute, University of Copenhagen, DK-2100 Copenhagen, Denmark \\
$^{22}$ Dept. of Physics, TU Dortmund University, D-44221 Dortmund, Germany \\
% ...
$^{35}$ Dept. of Physics and Astronomy, University of Kansas, Lawrence, KS 66045, USA \\
% ...
$^{39}$ Dept. of Physics and Wisconsin IceCube Particle Astrophysics Center, University of Wisconsin{\textemdash}Madison, Madison, WI 53706, USA \\
% ...
$^{41}$ Department of Physics, Marquette University, Milwaukee, WI 53201, USA \\
% ...
$^{43}$ Bartol Research Institute and Dept. of Physics and Astronomy, University of Delaware, Newark, DE 19716, USA \\
% ...
$^{63}$ Deutsches Elektronen-Synchrotron DESY, Platanenallee 6, D-15738 Zeuthen, Germany \\
$^{\rm a}$ also at Institute of Physics, Sachivalaya Marg, Sainik School Post, Bhubaneswar 751005, India \\
% ...

\subsection*{Acknowledgments}

\noindent
The authors gratefully acknowledge the support from the following agencies and institutions:
USA {\textendash} U.S. National Science Foundation-Office of Polar Programs,
U.S. National Science Foundation-Physics Division,
U.S. National Science Foundation-EPSCoR,
U.S. National Science Foundation-Office of Advanced Cyberinfrastructure,
Wisconsin Alumni Research Foundation,
Center for High Throughput Computing (CHTC) at the University of Wisconsin{\textendash}Madison,
Open Science Grid (OSG),
Partnership to Advance Throughput Computing (PATh),
Advanced Cyberinfrastructure Coordination Ecosystem: Services {\&} Support (ACCESS),
Frontera and Ranch computing project at the Texas Advanced Computing Center,
U.S. Department of Energy-National Energy Research Scientific Computing Center,
Particle astrophysics research computing center at the University of Maryland,
Institute for Cyber-Enabled Research at Michigan State University,
Astroparticle physics computational facility at Marquette University,
NVIDIA Corporation,
and Google Cloud Platform;
Belgium {\textendash} Funds for Scientific Research (FRS-FNRS and FWO),
FWO Odysseus and Big Science programmes,
and Belgian Federal Science Policy Office (Belspo);
Germany {\textendash} Bundesministerium f{\"u}r Forschung, Technologie und Raumfahrt (BMFTR),
Deutsche Forschungsgemeinschaft (DFG),
Helmholtz Alliance for Astroparticle Physics (HAP),
Initiative and Networking Fund of the Helmholtz Association,
Deutsches Elektronen Synchrotron (DESY),
and High Performance Computing cluster of the RWTH Aachen;
Sweden {\textendash} Swedish Research Council,
Swedish Polar Research Secretariat,
Swedish National Infrastructure for Computing (SNIC),
and Knut and Alice Wallenberg Foundation;
European Union {\textendash} EGI Advanced Computing for research;
Australia {\textendash} Australian Research Council;
Canada {\textendash} Natural Sciences and Engineering Research Council of Canada,
Calcul Qu{\'e}bec, Compute Ontario, Canada Foundation for Innovation, WestGrid, and Digital Research Alliance of Canada;
Denmark {\textendash} Villum Fonden, Carlsberg Foundation, and European Commission;
New Zealand {\textendash} Marsden Fund;
Japan {\textendash} Japan Society for Promotion of Science (JSPS)
and Institute for Global Prominent Research (IGPR) of Chiba University;
Korea {\textendash} National Research Foundation of Korea (NRF);
Switzerland {\textendash} Swiss National Science Foundation (SNSF).